\documentclass[%
 reprint,
superscriptaddress,
showpacs,preprintnumbers,
 amsmath,amssymb, 
 aps,
 notitlepage,
]{revtex4-1}

\usepackage{graphicx}% Include figure files
\usepackage{dcolumn}% Align table columns on decimal point
\usepackage{bm}% bold math
\usepackage{float}% float
\usepackage{hyperref}% add hypertext capabilities
\begin{document}

\title{\boldmath{Evidence for $e^{+}e^{-} \to \gamma\eta_c(1S)$ at center-of-mass energies between 4.01 and 4.60 GeV}}

\author{M.~Ablikim$^{1}$, M.~N.~Achasov$^{9,d}$, S. ~Ahmed$^{14}$, M.~Albrecht$^{4}$, M.~Alekseev$^{53A,53C}$, A.~Amoroso$^{53A,53C}$, F.~F.~An$^{1}$, Q.~An$^{50,40}$, J.~Z.~Bai$^{1}$, Y.~Bai$^{39}$
% ...
}

\affiliation{
Institute of High Energy Physics, Beijing 100049, People's Republic of China\\
% ...
$^{4}$ Bochum Ruhr-University, D-44780 Bochum, Germany\\
% ...
$^{9}$ G.I. Budker Institute of Nuclear Physics SB RAS (BINP), Novosibirsk 630090, Russia\\
% ...
$^{14}$ Helmholtz Institute Mainz, Johann-Joachim-Becher-Weg 45, D-55099 Mainz, Germany\\
% ...
$^{39}$ Southeast University, Nanjing 211100, People's Republic of China\\
$^{40}$ State Key Laboratory of Particle Detection and Electronics, Beijing 100049, Hefei 230026, People's Republic of China\\
% ...
$^{50}$ University of Science and Technology of China, Hefei 230026, People's Republic of China\\
% ...
$^{53}$ (A)University of Turin, I-10125, Turin, Italy; (B)University of Eastern Piedmont, I-15121, Alessandria, Italy; (C)INFN, I-10125, Turin, Italy\\
% ...
$^{d}$ Also at the Novosibirsk State University, Novosibirsk, 630090, Russia\\
% ...
}
 \collaboration{BESIII Collaboration}
 \noaffiliation

\begin{abstract}
We present first evidence for the process $e^+e^-\to \gamma\eta_c(1S)$ at six center-of-mass energies between 4.01 and 4.60~GeV using data collected by the BESIII experiment operating at BEPCII. 
These data sets correspond to a total integrated luminosity of 4.6~fb$^{-1}$. 
We measure the Born cross section at each energy using a combination of twelve $\eta_c(1S)$ decay channels.  
Because the significance of the signal is marginal at each energy ($\le3.0\sigma$), we also combine all six energies under various assumptions for the energy-dependence of the cross section. 
If the process is assumed to proceed via the $Y(4260)$, we measure a peak Born cross section $\sigma_{\mathrm{ peak}}(e^+e^-\to\gamma\eta_c(1S)) = 2.11 \pm 0.49 (\mathrm{stat.}) \pm 0.36 (\mathrm{syst.})$~pb with a statistical significance of 4.2$\sigma$. 
\end{abstract}

\maketitle
%\tableofcontents

The $Y(4260)$, first discovered by BaBar in the initial state radiation~(ISR) process $e^+e^-\to \gamma_{\mathrm{ISR}} Y(4260) \to  \gamma_{\mathrm{ISR}} \pi^+\pi^-J/\psi$~\cite{BaBarY4260Discovery}, cannot be easily explained within the traditional $c\bar{c}$ picture of charmonium.  
From its production mechanism, we know its spin~($J$), parity~($P$), and charge-parity~($C$) quantum numbers are $J^{PC} = 1^{--}$.  However, due to its distinct mass, it cannot be identified with the previously established $\psi$ states in this region~\cite{PDG}.  Furthermore, while the $\psi(4040)$, $\psi(4160)$, and $\psi(4415)$ states are thought to be the $n^{2S+1}L_J$ = $3^3S_1$, $2^3D_1$, and $4^3S_1$ states of charmonium, respectively~\cite{TheoryGammaEtac1}, the $Y(4260)$ appears to be supernumerary. 

One possibility is that the $Y(4260)$ is a hybrid meson \cite{YHybrid1,YHybrid2}.  If so, recent lattice QCD calculations predict that its rate of decay to $\gamma\eta_c(1S)$ will be enhanced relative to $\gamma\chi_{c0}(1P)$~\cite{Dudek2009}.  This is in stark contrast to the pattern for conventional $\psi$ states, where, for example, the $\psi(2S)$ decays to $\gamma\chi_{c0}(1P)$ about 30 times more often than to $\gamma\eta_c(1S)$.  Finding evidence for $Y(4260) \to \gamma \eta_c(1S)$ could thus give additional support to the hybrid interpretation.  

In this paper, we search for the process $e^+e^- \to \gamma \eta_c$ (where $\eta_c$ always denotes $\eta_c(1S)$) using data collected by the BESIII detector operating at the Beijing Electron Positron Collider~(BEPCII). 
We use a total integrated luminosity of 4.6~fb$^{-1}$ spread among six center-of-mass energies ($E_{CM}$): 482~pb$^{-1}$ at 4.01~GeV,  1092~pb$^{-1}$ at 4.23~GeV,  826~pb$^{-1}$ at 4.26~GeV,  540~pb$^{-1}$ at 4.36~GeV,  1074~pb$^{-1}$ at 4.42~GeV,  and 567~pb$^{-1}$ at 4.60~GeV~\cite{Ablikim,BesIIIBeamEnergies}.  

We first measure the Born cross section at each $E_{CM}$ using the twelve largest decay channels of the $\eta_c$: 
$2(\pi^{+}  \pi^{-} \pi^{0})$, 
$\pi^{+}  \pi^{-}  \pi^{0}  \pi^{0}$, 
$\pi^{+}  \pi^{+}  \pi^{-}  \pi^{-}  \eta$, 
$K^{+}  K^{-}  \pi^{+}  \pi^{-}  \pi^{0}$, 
$2(\pi^{+}  \pi^{-})$, 
$3(\pi^{+}  \pi^{-})$, 
$\pi^{+}  \pi^{-}  \eta$, 
$K^{\pm}  K_{S}  \pi^{\mp}  \pi^{+}  \pi^{-}$, 
$K^{\pm}  K_{S}  \pi^{\mp}$, 
$K^{+}  K^{-}  \pi^{0}$, 
$K^{+}  K^{-}  \pi^{+}  \pi^{-}$, 
and $K^{+}  K^{-}  \pi^{+}  \pi^{+}  \pi^{-}  \pi^{-}$.
We then combine the data from the six $E_{CM}$ under four different assumptions about the energy-dependence of the cross section:
(1)~$\sigma_{\mathrm{FLAT}}$: the cross section is constant, consistent with the calculation in Ref.~\cite{NonResonant};
(2)~$\sigma_{\mathrm{BELLE}}$: the cross section follows the Belle parameterization of $\sigma(e^+e^-\to\pi^+\pi^-J/\psi)$ found in Ref.~\cite{CombinedPsi1SBelle}, modeled with a $Y(4008)$ in addition to the $Y(4260)$;
(3)~$\sigma_{\mathrm{Y(4260)}}$: the cross section follows a non-relativistic Breit-Wigner distribution for the $Y(4260)$ with mass and width values from the Particle Data Group (PDG)~\cite{PDG};
and (4)~$\sigma_{\mathrm{Y(4360)}}$:  the cross section follows a non-relativistic Breit-Wigner distribution for the $Y(4360)$ with mass and width values from the PDG.
Combining the data samples in this way allows us to search for $e^+e^-\to\gamma\eta_c$ using a larger sample of events and 
allows us to compare the $Y(4260)$ hypothesis~($\sigma_{\mathrm{Y(4260)}}$) to other hypotheses.

The BEPCII $e^+e^-$ storage ring is designed to have a peak luminosity of $10^{33}~\mathrm{cm}^{-2}\mathrm{s}^{-1}$ at a beam energy of 1.89~GeV~\cite{Asner2009}. The BESIII detector is a general purpose hadron detector built around the collision point at BEPCII~\cite{BESIIIdesign}. Charged particles are detected in the main drift chamber~(MDC) and are bent by an on-axis 1~Tesla solenoidal magnetic field, yielding a momentum resolution of 0.5\% at 1~GeV/$c$.
Time-of-flight~(TOF) scintillation counters are placed around the MDC and provide a timing resolution of 80~ps in the barrel and 110~ps in the end caps. 
Photons are detected by the Electromagnetic Calorimeter~(EMC) surrounding the TOF. 
The photon energy resolution at 1~GeV is 2.5\% in the barrel and 5\% in the end caps. 
The geometric acceptance is 93\% of 4$\pi$.

The response of the BESIII detector is modeled using Monte Carlo~(MC) simulation software based on \textsc{geant4}~\cite{GEANT4}.  To study signal efficiencies, mass resolutions, cross-feeds among $\eta_c$ decay channels, and effects due to ISR, a series of MC data samples were generated according to the signal process $e^+e^-\to\gamma\eta_c$, where the $\eta_c$ subsequently decays to the twelve channels listed above.  ISR effects are modeled using \textsc{kkmc}~\cite{KKMC1,KKMC2}.  The production of $\gamma\eta_c$ and the subsequent decays of the $\eta_c$ are handled by \textsc{evtgen}~\cite{EvtGen,KKMC3} using kinematics following phase space distributions.  To study background processes, we generate large samples of generic $q\bar{q}$ events as well as samples corresponding to the ISR process $e^+e^-\to \gamma_{\mathrm{ISR}} J/\psi$, where the $J/\psi$ either decays to the same twelve modes as the $\eta_c$ or decays to $\gamma\eta_c$.

We reconstruct events of the form $\gamma X_i$, where the $\gamma$ is referred to as the ``transition photon'' and the $X_i$ are the twelve different combinations of hadrons corresponding to the $\eta_c$ decay channels listed above.
The criteria used to select events have been optimized using both MC samples and sidebands of the $\eta_c$ from data. 

Charged pions and kaons are reconstructed using information from the MDC.  Their angle with respect to the beam direction, $\theta$, must satisfy $\lvert\cos\theta\rvert<0.93$.  Except for pions originating from $K_S$ decays, all charged tracks are further required to pass within 10~cm of the interaction point along the beam direction and within 1~cm in a plane perpendicular to the beam.  Pions (except for pions originating from $K_S$ decays) and kaons are separated using a combination of ionization energy loss in the MDC and timing information from the TOF.  For each reconstructed track, particle identification probabilities $P_\pi$ and $P_K$ are calculated based on pion and kaon hypotheses, respectively.  For pions, we require $P_\pi > 10^{-5}$; for kaons, we require $P_K > 10^{-5}$ and $P_K > P_\pi$. 

Photons are reconstructed in the EMC by clustering energies deposited in individual crystals.
Energy clusters in the barrel region ($|\cos\theta|<0.8$) must be greater than 25~MeV and they must be greater than~50 MeV in the end cap region ($0.86<|\cos\theta|<0.92$). Timing from the EMC is used to suppress electronic noise and background from unrelated events.  We reject candidate transition photons that can be paired with any other energy cluster in an event to form a $\pi^0$.  In the $\pi^+\pi^-\eta$ channel, the candidate transition photon is isolated from clusters formed by charged tracks by requiring their angle of separation be greater than 17.5$^{\circ}$.

%\begin{figure}[t]
%\includegraphics[width = 1.0\columnwidth]{figures/fig1.pdf}
%\caption{\label{fig:indiEnergy} 
%The recoil-mass distribution of the transition photon for each $\eta_c$ decay channel at $E_{CM}=4.23$~GeV.  Projections from the simultaneous fit are overlaid.
%Dotted and dashed vertical lines indicate the $\eta_c$ and $J/\psi$ masses, respectively. }
%\end{figure}

We form $\pi^0$ and $\eta$ candidates using combinations of two photons with invariant mass satisfying $107<M(\gamma\gamma)<163$~MeV/$c^2$ and $400<M(\gamma\gamma)<700$~MeV/$c^2$, respectively.  Similarly, $K_S$ candidates are formed using two oppositely charged tracks, assumed to be pions, satisfying $471<M(\pi^+\pi^-)<524$~MeV/$c^2$.

From these initial lists of $\gamma$, $\pi^\pm$, $K^\pm$, $\pi^0$, $\eta$, and $K_S$, we form all possible combinations of $\gamma X_i$ for each $i$.  We perform a kinematic fit for each of these combinations to the initial four-momentum of the center-of-mass system~(4C) and add one constraint~(1C) for the mass of every $\pi^0$, $\eta$, and $K_S$ candidate.  We require that the resulting $\chi^2$ per degree of freedom (dof) be less than a value optimized separately for each $X_i$, ranging from $3.0$ to $5.2$. 
To avoid multiple counting, we only use the combination with the best $\chi^2$/dof.
Reconstruction efficiencies after all event selection range from 4\% ($\eta_c\to\pi^+\pi^+\pi^-\pi^-\pi^0\pi^0$) to 35\% ($\eta_c\to\pi^+\pi^+\pi^-\pi^-$).

% fits at each energy

\begin{figure}[t]                          
\centerline{
\includegraphics[width = 0.9\columnwidth]{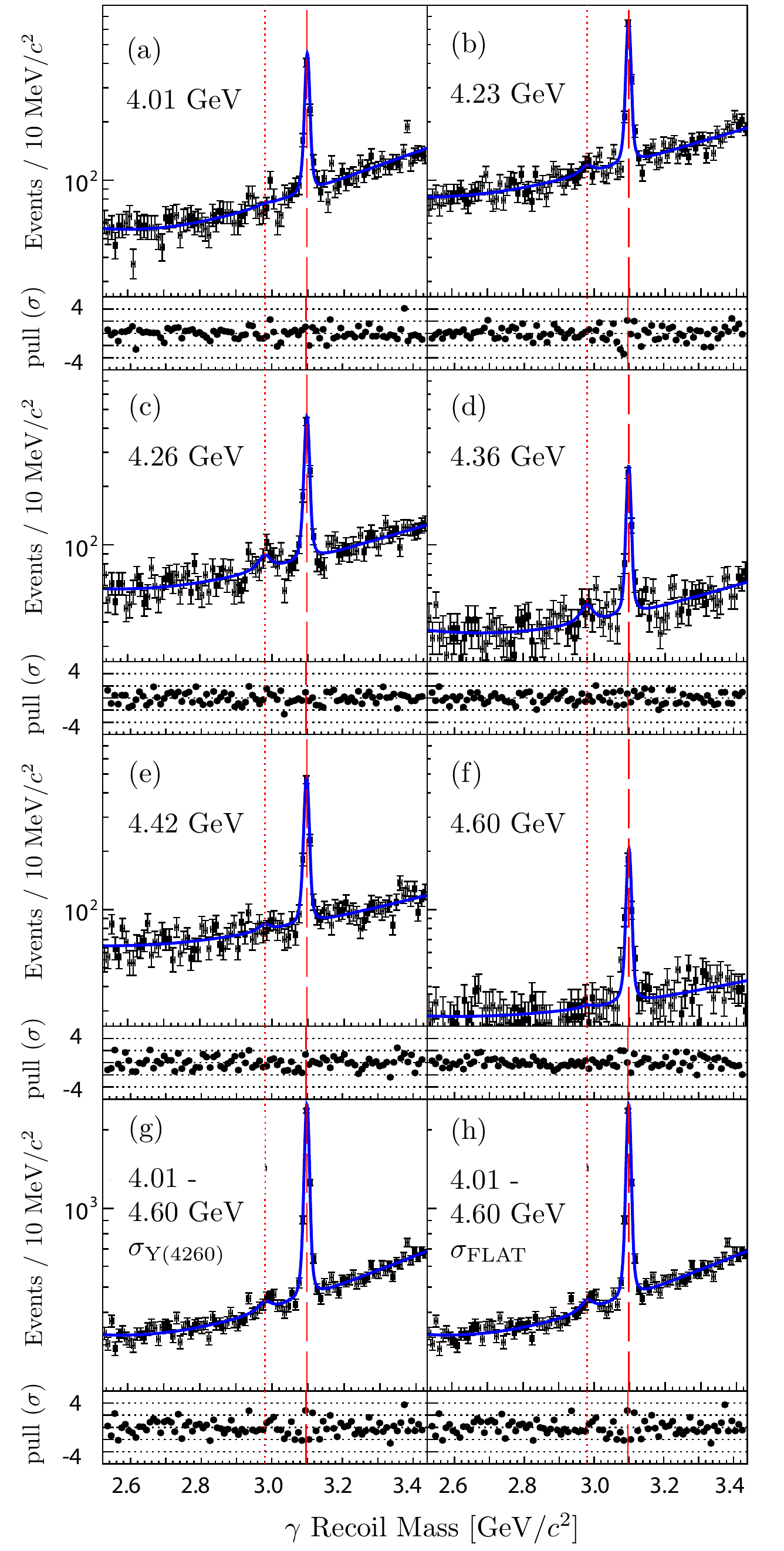}}
\caption{\label{fig:combEnergy} 
The recoil-mass distribution of the transition photon summed over all $\eta_c$ decay channels.  Results from the simultaneous fits are overlaid.  In (a-f) the fits are performed separately at each energy; in (g) and (h) the data are combined and fit with the $\sigma_{\mathrm{Y(4260)}}$ and $\sigma_{\mathrm{FLAT}}$ assumptions, respectively.  Pull distributions, derived by comparing the fit projections and the data, are shown below each plot.  Dotted and dashed vertical lines indicate the $\eta_c$ and $J/\psi$ masses, respectively.}
\end{figure}

To determine the Born cross section at each $E_{CM}$, 
we use an unbinned maximum likelihood method to simultaneously fit the recoil-mass distributions of the transition photon associated with the twelve final states $\gamma X_i$.  
%The fit at 4.23~GeV is shown in Fig.~\ref{fig:indiEnergy}.  
The total fit projections from each of the six $E_{CM}$ are shown in Fig.~\ref{fig:combEnergy}(a-f).
The fit range is centered at the $\eta_c$ mass and extends 450~MeV/$c^2$ on either side.  The $\eta_c$ signal is described by a non-relativistic Breit-Wigner function convolved with a histogram derived from MC describing detector resolution and effects due to ISR.  The mass and width of the $\eta_c$ are fixed to their PDG values.  The Born cross section, $\sigma(e^+e^-\to\gamma\eta_c)$, is a shared free parameter that accounts for $\eta_c$ decay branching fractions, reconstruction efficiencies, corrections due to ISR effects~\cite{BudkerISR,ISRCorrection} (evaluated using the $\sigma_{\mathrm{Y(4260)}}$ assumption), vacuum polarization~\cite{VacPolarization}, and integrated luminosity.

\begin{figure}[t]
\includegraphics[width = 0.8\columnwidth]{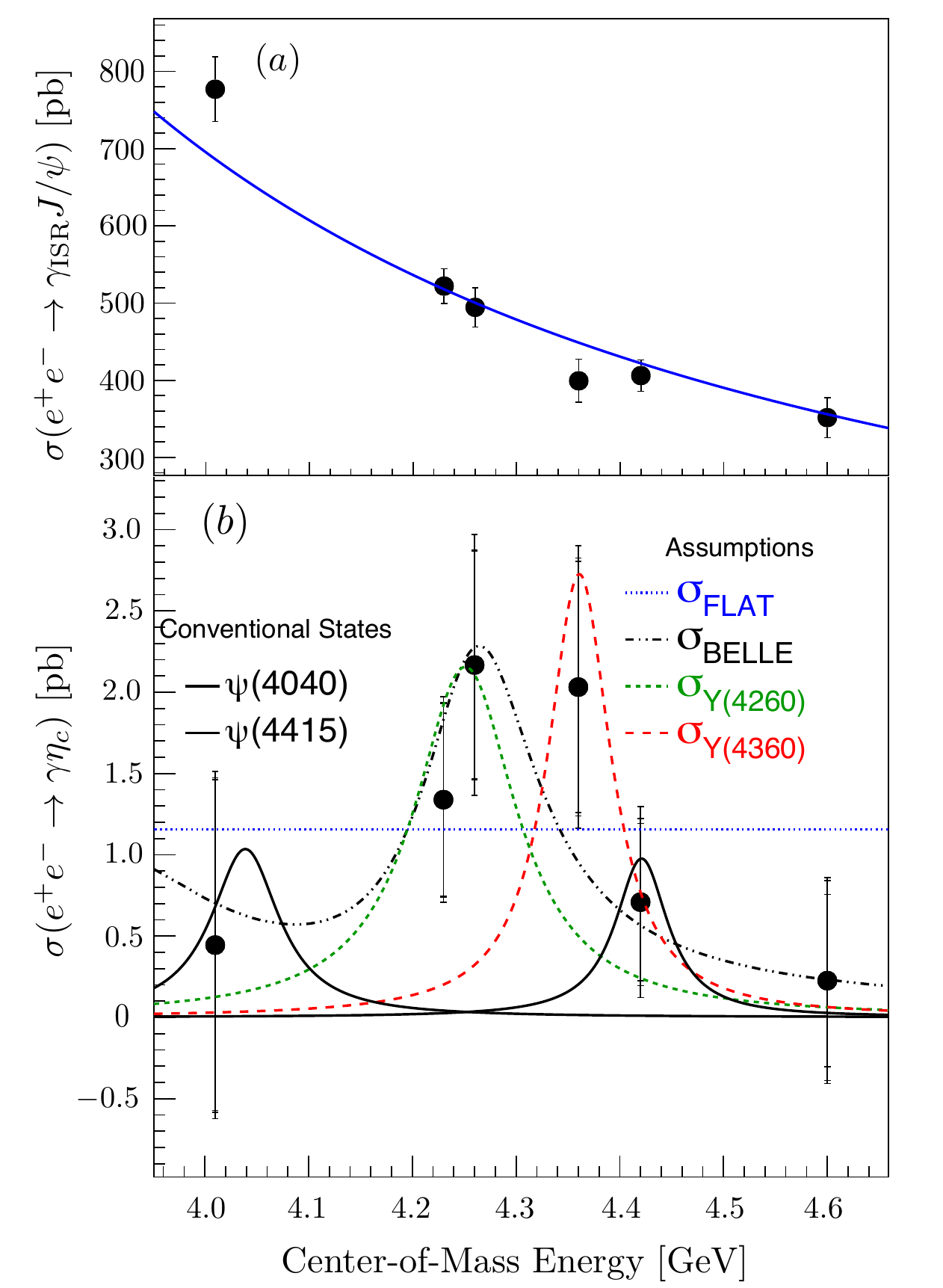}
\caption{\label{fig:overlay}
(a)~The cross section for $e^+e^-\to \gamma_{\mathrm{ISR}} J/\psi$~(points) compared to the theoretical calculation~
(line)~\cite{BudkerISR,CLEOJPsiISR}. 
(b)~The Born cross section for $e^+e^-\to\gamma\eta_c$
measured at each $E_{CM}$~(points) and measured using the sum of all the data under various assumptions about the energy-dependence of the cross section~(broken lines).  The first tick marks are due to the statistical uncertainty, the intermediate tick marks sum in quadrature the statistical and the systematic uncertainties uncorrelated in energy (see Table~\ref{tab:table1}), and the outermost tick marks sum in quadrature both the statistical and total systematic uncertainties.  The predicted cross sections for $e^+e^-\to\psi(4040)\to\gamma\eta_c$ and $e^+e^-\to\psi(4415)\to\gamma\eta_c$~\cite{TheoryGammaEtac1} are shown as solid lines.}
\end{figure}

The major backgrounds in the recoil-mass distribution of the transition photon are from the continuum $q\bar{q}$ process and the $J/\psi$ ISR process, $e^+e^-\to \gamma_{\mathrm{ISR}} J/\psi$, where the $J/\psi$ decays to the same channels as the $\eta_c$.  The potential background where the $J/\psi$ decays to $\gamma\eta_c$ has been found to be negligible.  The continuum background is described independently in each decay channel using a second order polynomial function. The peaking $J/\psi$ ISR background is parameterized by a double Gaussian function whose parameters are fixed using MC studies. The size of the $J/\psi$ ISR background is allowed to float independently in each decay channel.

Since the $J/\psi$ ISR cross section, $\sigma(e^+e^- \to \gamma_{\mathrm{ISR}} J/\psi)$, can be accurately calculated using a combination of the ISR rate~\cite{BudkerISR} and $\sigma(e^+e^-\to J/\psi)$~\cite{CLEOJPsiISR}, this process serves as an important cross-check to the $\eta_c$ analysis.  When we perform a simultaneous fit that constrains the size of the $J/\psi$ ISR background among the $X_i$ using known $J/\psi$ decay branching fractions, we obtain the results shown in Fig.~\ref{fig:overlay}(a).  There is good agreement between the measurements and the theoretical predictions.  
We also obtain good agreement with the average $J/\psi$ cross section when the size of the $J/\psi$ ISR background is not constrained among the $X_i$, although with less precision.

Our final measurements of $\sigma(e^+e^-\to\gamma\eta_c)$ are listed in Table~\ref{tab:finalObservedXC} and are shown as the points in Fig.~\ref{fig:overlay}(b).  
These use the $\sigma_{\mathrm{Y(4260)}}$ assumption for the calculation of effects due to ISR.  The other assumptions are also used and the differences range from 1\% to 6\%, which are included in the systematic uncertainties.
Significances of the $\eta_c$ signal are obtained by comparing the likelihoods of fits with and without the $\eta_c$ signal.  The largest significance~(3.0$\sigma$) is found at $E_{CM}$ = 4.26~GeV. 
Upper limits of the Born cross section (at 90\% confidence level) are calculated by first convolving the likelihood function with a Gaussian function whose width corresponds to the total systematic uncertainty, then integrating the resulting likelihood function up to the value that includes 90\% of the integral.

\begin{table}[ht]
\caption{\label{tab:finalObservedXC} 
Measurements of the Born cross section $\sigma(e^+e^-\to\gamma\eta_c)$ (where the first uncertainty is statistical and the second is systematic), statistical significance~(sig.), and 90\% confidence level upper limits (U.L.) at each $E_{CM}$.
}
\begin{ruledtabular}
\begin{tabular}{cccc}
$E_{CM}$ (GeV)  &  $\sigma(e^+e^-\to\gamma\eta_c)$ (pb) &  sig. ($\sigma$)  &  U.L. (pb)  \\
\hline
4.01  &  0.44 $\pm$ 1.02 $\pm$ 0.32  &  0.4  &	 2.4 \\
4.23  &  1.34 $\pm$ 0.59 $\pm$ 0.22  &  2.2  &	 2.2 \\
4.26  &  2.17 $\pm$ 0.70 $\pm$ 0.39  &  3.0  &	 3.2 \\
4.36  &  2.03 $\pm$ 0.77 $\pm$ 0.40  &  2.7  &	 3.2 \\
4.42  &  0.71 $\pm$ 0.48 $\pm$ 0.33  &  1.4  &	 1.6 \\
4.60  &  0.23 $\pm$ 0.53 $\pm$ 0.35  &  0.4  &	 1.4 \\
\end{tabular}
\end{ruledtabular}
\end{table}

\begin{table}[ht]
\caption{\label{tab:MultiEnergyBornXC} Measurements of the peak Born cross section $\sigma_{\mathrm{peak}}(e^{+}e^{-}\rightarrow\gamma\eta_{c})$ under various assumptions for the energy-dependence of the cross section.
}
\begin{ruledtabular}
\begin{tabular}{cccc}
assumption  &  $\sigma_{\mathrm{peak}}(e^{+}e^{-}\rightarrow\gamma\eta_{c})$ (pb)  &  sig. ($\sigma$)   & U.L. (pb)  \\
\hline
$\sigma_{\mathrm{FLAT}}$  &  1.16 $\pm$ 0.27 $\pm$ 0.20  &  4.1     &  1.6  \\
$\sigma_{\mathrm{BELLE}}$  &  2.27 $\pm$ 0.49 $\pm$ 0.39  &  4.5    &  3.1  \\
$\sigma_{\mathrm{Y(4260)}}$  &  2.11 $\pm$ 0.49 $\pm$ 0.36  &  4.2  &  2.9  \\
$\sigma_{\mathrm{Y(4360)}}$  &  2.72 $\pm$ 0.71 $\pm$ 0.46  &  3.6  &  3.9   \\
\end{tabular}
\end{ruledtabular}
\end{table}

% combined fits

Because there is little evidence for the $e^+e^-\to\gamma\eta_c$ process at any individual energy, we combine all six energies under various assumptions for the energy-dependence of the cross section. In this case, we perform a simultaneous fit to the $6\times 12$ recoil-mass distributions of the transition photon.  At each energy, the $\gamma\eta_c$ cross section is constrained to be the same, as before.  But between the different energies, the cross section is now constrained to follow the $\sigma_{\mathrm{FLAT}}$, $\sigma_{\mathrm{BELLE}}$, $\sigma_{\mathrm{Y(4260)}}$, or $\sigma_{\mathrm{Y(4360)}}$
cross section assumptions. 
Table~\ref{tab:MultiEnergyBornXC} lists the final peak cross sections using this method, where the peak is measured at 4.26~GeV for the $\sigma_{\mathrm{Y(4260)}}$ and $\sigma_{\mathrm{BELLE}}$ assumptions, and at 4.36~GeV for $\sigma_{\mathrm{Y(4360)}}$. The statistical significances of the $\eta_c$ signal and the upper limits on the Born cross sections are determined as before.  Figure~\ref{fig:combEnergy}(g-h) shows no observable difference in the fit projections for the $\sigma_{\mathrm{Y(4260)}}$ and $\sigma_{\mathrm{FLAT}}$
assumptions. The lines in Fig.~\ref{fig:overlay}(b) show the resulting cross sections as a function of energy.
The statistical significance of the $\gamma\eta_c$ process is at least 3.6$\sigma$, regardless of our input cross section assumption.

While we find evidence for $e^+e^-\to\gamma\eta_c$ in our combined fits, we are unable to distinguish between the different assumptions for the energy dependence of the cross section.  To test the significance of the $\sigma_{\mathrm{Y(4260)}}$ shape, we compare the likelihood value of a fit assuming a combination of $\sigma_{\mathrm{Y(4260)}}$ and $\sigma_{\mathrm{FLAT}}$ (where the sizes of both components are free parameters in the fit) to that of the fit assuming $\sigma_{\mathrm{FLAT}}$.  In this test, we find the significance of the $\sigma_{\mathrm{Y(4260)}}$ component to be only 1.5$\sigma$.

The expected rate of $e^+e^-\to \psi(4040) \to \gamma \eta_c$, shown as a solid line in Fig.~\ref{fig:overlay}(b), is estimated using the calculated partial width $\Gamma(\psi(4040)\to\gamma \eta_c)$~\cite{TheoryGammaEtac1}. If we assume the energy-dependence of the cross section follows the $\psi(4040)$ and fit our combined data sets allowing the size of the $\psi(4040)$ to float, then the significance of $e^+e^-\to\gamma\eta_c$ is 1.9$\sigma$. Predictions of $\psi(4160)$ or $\psi(4415)$ to $\gamma\eta_c$ have not been published but are calculable using the models discussed in~\cite{TheoryGammaEtac1}. 
The expected rate of $e^+e^-\to \psi(4415) \to \gamma \eta_c$ is also shown as a solid line in Fig.~\ref{fig:overlay}(b). The significance of $e^+e^-\to\gamma\eta_c$ assuming the $\psi(4415)$ is 1.9$\sigma$. In the case of the $\psi(4160)$ we are missing crucial data at $E_{CM}$ near 4.16~GeV to constrain this assumption. Nevertheless, we measure the significance of $e^+e^-\to\gamma\eta_c$ assuming $\psi(4160)$ production 
to be 3.5$\sigma$, 
which is still less significant than all other nonconventional assumptions.

% systematic errors

Estimates of the systematic uncertainty on the cross section measurements, discussed individually below, are summarized in Table~\ref{tab:table1}.
The total systematic uncertainty is obtained by adding the individual systematic uncertainties in quadrature.

\begin{table}[ht]
\caption{\label{tab:table1}
Systematic errors (in percent) on the cross section measured at each $E_{CM}$ and for all $E_{CM}$ combined~(All).
Errors with an asterisks (*) are correlated among $E_{CM}$.
}
\begin{ruledtabular}
\begin{tabular}{cccccccc}
  $E_{CM}$ (GeV) &  4.01  &  4.23  &  4.26  &  4.36  &  4.42  &  4.60  &  All  \\
  \hline
  * $\mathcal{B}(\eta_{c}\to X_i)$  &  41     &  9   &  12   &  11   &  18    &  38     &  7   \\
  MC statistics                     &  2      &  1   &  1    &  1    &  1     &  2      &  1   \\
  * Mass resolution                 &  43     &  6   &  8    &  6    &  17    &  42     &  10   \\
  * $\eta_{c}$ mass and width       &  10     &  1   &  2    &  3    &  3     &  3      &  1   \\
  $e^{+}e^{-}$ beam energy          &  7      &  1   &  1    &  2    &  1     &  3      &  1   \\
  * $\eta_c$ lineshape              &  4      &  7   &  1    &  5    &  30    &  31     &  3   \\
  * Tracking efficiency             &  16     &  7   &  9    &  9    &  8     &  12     &  8   \\
  * Photon efficiency               &  2      &  3   &  4    &  3    &  4     &  4      &  3   \\
  * $K_{S}$ efficiency              &  2      &  1   &  2    &  1    &  1     &  3      &  4   \\
  * Kinematic fitting               &  5      &  1   &  1    &  3    &  2     &  2      &  2   \\
  Background Shape                  &  29     &  4   &  2    &  7    &  23    &  123    &  5   \\
  $J/\psi$ peak                     &  20     &  4   &  1    &  1    &  7     &  62     &  2   \\
  $\sigma_E$ assumption   &  2      &  2   &  3    &  5    &  3     &  6      &    \\
  Luminosity                        &  1      &  1   &  1    &  1    &  1     &  1      &  1   \\
  \hline
  Total  &  73   &  16   &  18   &  20   &  47   &  153   &  17   \\
\end{tabular}
\end{ruledtabular}
\end{table}

One of the largest systematic uncertainties comes from uncertainty in the branching fractions of the $\eta_c$ decays.  We estimate this uncertainty by performing many trials of our simultaneous fitting procedure using different input $\eta_c$ branching fractions,
which are randomly generated according to their uncertainties.
When available, we use the branching fractions measured by BESIII in Ref.~\cite{Ablikim2012}.
Since those measurements were performed by taking the ratio of $\mathcal{B}(\psi(2S)\to\pi^0h_c(1P))\times \mathcal{B}(h_c(1P)\to\gamma\eta_c)\times \mathcal{B}(\eta_c\to X_i))$ with $\mathcal{B}(\psi^\prime\to\pi^0h_c(1P))\times \mathcal{B}(h_c(1P)\to\gamma\eta_c)$, 
we account for correlated errors by first randomly varying the denominator (the double product), then varying the numerator (the triple product) 
for each $X_i$, and derive $\eta_c$ branching fractions using the common denominator. The RMS of the resulting $e^+e^-\to\gamma\eta_c$ cross sections are taken as the systematic uncertainty. Note that the $\eta_c$ branching fraction measurements include systematic uncertainties due to the substructure in $\eta_c$ decays. 

We estimate the uncertainty due to the statistical uncertainty of the efficiencies (ranging from 1 to 2\%) using the same procedure.  That is, we perform many trials of the fits while varying the efficiencies according to their uncertainties.

In our baseline fits to the recoil-mass distribution of the transition photon, we use a resolution derived from MC for both the $\eta_c$ signal and the $J/\psi$ ISR background. By studying the $J/\psi$ ISR peak in its largest decay channels, we have found the resolution in data is wider than that in MC by up to 20\%.  We estimate the systematic uncertainty that this introduces by repeating the fits with a resolution widened by a factor of 1.2.

To estimate the uncertainty caused by fixing the $\eta_c$ mass and width to their PDG averages, we vary them by $\pm1\sigma$, repeat the fits, and take the largest difference as a systematic uncertainty. Our nominal values of the $E_{CM}$ are taken from Ref.~\cite{BesIIIBeamEnergies}, but an uncertainty in the $E_{CM}$ can cause a 0.75~MeV/$c^2$ shift in the apparent mass of the $\eta_c$. We also vary the input $\eta_c$ mass by $\pm$0.75~MeV/$c^2$ to account for this possibility. 

To account for a possible distortion in the $\eta_c$ signal shape due to the photon energy-dependence of electromagnetic transitions~\cite{Mitchell2009,Anashin2010}, we repeat the fit using the $\eta_c$ signal shape developed
in Ref.~\cite{Anashin2010}.

We assign an uncertainty of 2\% per charged pion and kaon to account for uncertainty in the track reconstruction efficiency~(including particle~ID)~\cite{TrackSys1,TrackSys2}.  The error due to uncertainty in photon reconstruction efficiencies is 1\% per photon (including photons from $\pi^0$ and $\eta$)~\cite{PhotonSystematic}.  The total error attributed to the $K_{S}$ reconstruction efficiency (arising from a combination of geometric acceptance, tracking efficiency, and selection efficiency) is 4\% per $K_{S}$~\cite{KsSystematic}. 
We vary the efficiency in each $\eta_c$ channel by its positive and negative extremes, refit data, and take the largest difference with respect to the nominal measurement as the systematic uncertainty.

Uncertainties in the kinematic fitting efficiencies are evaluated by comparing the cross sections extracted with and without tracking corrections, following the method used in Ref.~\cite{Yuping2013}.

To judge our sensitivity to the background shape, we try a third order polynomial function in place of the second order polynomial function used in the baseline fits.  We take the difference as a systematic uncertainty.

In the baseline fits, the size of the $J/\psi$ peak is allowed to float independently in each channel.  We also fix the relative size of the $J/\psi$ peak among channels using known $J/\psi$ branching fractions and take the difference as a systematic uncertainty.

In summary, we search for the process $e^+e^- \to \gamma\eta_c$ at six $E_{CM}$ between 4.01 and 4.60~GeV using 4.6~fb$^{-1}$ of data collected by BESIII.  
While we do not find evidence for this process at any individual energy, the significance is consistently above 3$\sigma$ when we combine all of our data sets according to the four assumptions listed above.  With our current statistics, we cannot make firm conclusions about the energy-dependence of the cross section. 
However, we note that the cross section is better explained by $\sigma_{\mathrm{Y(4260)}}$ than by conventional charmonium states: $\psi(4040)$, $\psi(4160)$, and $\psi(4415)$. 
Although we are unable to unambiguously determine the production mechanism of $\gamma\eta_c$, the enhancement in $e^+e^-\to\gamma\eta_c$ between 4.23 and 4.36~GeV may suggest production via a hybrid charmonium state.
Measurements of cross sections for other reactions, especially for $e^+e^- \to \gamma \chi_{c0}(1P)$, are required to make further progress.

If we assume $e^+e^-\to\gamma \eta_c$ proceeds through a $Y(4260)$, we measure $\sigma_{\mathrm{peak}}(e^+e^-\to\gamma\eta_c) = 2.11 \pm 0.49 (\mathrm{stat.}) \pm 0.36 (\mathrm{syst.})$~pb.  Combining this with a previous BESIII measurement of $\sigma(e^+e^-\to \pi^+\pi^- J/\psi)$~\cite{BESIIIObservationY4260} at 4.26~GeV, 
we estimate $\mathcal{B}(Y(4260)\to\gamma\eta_c)/\mathcal{B}(Y(4260)\to\pi^+\pi^-J/\psi)$ = $0.034\pm0.009$, where the statistical and systematic uncertainties have been combined.

The authors would like to thank Eric Swanson for useful discussions about conventional charmonium decays to $\gamma\eta_c$. 
The BESIII collaboration thanks the staff of BEPCII and the IHEP computing center for their strong support. This work is supported in part by National Key Basic Research Program of China under Contract No. 2015CB856700; National Natural Science Foundation of China (NSFC) under Contracts Nos. 11235011, 11335008, 11425524, 11625523, 11635010; the Chinese Academy of Sciences (CAS) Large-Scale Scientific Facility Program; the CAS Center for Excellence in Particle Physics (CCEPP); Joint Large-Scale Scientific Facility Funds of the NSFC and CAS under Contracts Nos. U1332201, U1532257, U1532258; CAS under Contracts Nos. KJCX2-YW-N29, KJCX2-YW-N45, QYZDJ-SSW-SLH003; 100 Talents Program of CAS; National 1000 Talents Program of China; INPAC and Shanghai Key Laboratory for Particle Physics and Cosmology; German Research Foundation DFG under Contracts Nos. Collaborative Research Center CRC 1044, FOR 2359; Istituto Nazionale di Fisica Nucleare, Italy; Joint Large-Scale Scientific Facility Funds of the NSFC and CAS; Koninklijke Nederlandse Akademie van Wetenschappen (KNAW) under Contract No. 530-4CDP03; Ministry of Development of Turkey under Contract No. DPT2006K-120470; National Natural Science Foundation of China (NSFC); National Science and Technology fund; The Swedish Resarch Council; U. S. Department of Energy under Contracts Nos. DE-FG02-05ER41374, DE-SC-0010504, DE-SC-0010118, DE-SC-0012069; University of Groningen (RuG) and the Helmholtzzentrum fuer Schwerionenforschung GmbH (GSI), Darmstadt; WCU Program of National Research Foundation of Korea under Contract No. R32-2008-000-10155-0

\bibliography{etac}

\end{document}